\documentclass{article}

\pdfoutput=1
\usepackage{arxiv}

\usepackage[utf8]{inputenc} % allow utf-8 input
\usepackage[T1]{fontenc}    % use 8-bit T1 fonts
\usepackage{hyperref}       % hyperlinks
\usepackage{url}            % simple URL typesetting
\usepackage{booktabs}       % professional-quality tables
\usepackage{amsfonts}       % blackboard math symbols
\usepackage{nicefrac}       % compact symbols for 1/2, etc.
\usepackage{microtype}      % microtypography
\usepackage{lipsum}
\usepackage{float}
\usepackage{xcolor}

\usepackage[T1]{fontenc}
\usepackage[utf8]{inputenc}

\usepackage{graphicx}
\usepackage{multirow}
\begin{document}

\title{Detector-SegMentor Network for Skin Lesion Localization and Segmentation}

\author{
 Shreshth Saini \\
  Department of Electrical Engineering\\
  Indian Institute of Technology Jodhpur\\
  Jodhpur, India 342037 \\
  \texttt{saini.2@iitj.ac.in} \\
   \And
 Divij Gupta \\
  Department of Electrical Engineering\\
  Indian Institute of Technology Jodhpur\\
  Jodhpur, India 342037 \\
   \And
 Anil Kumar Tiwari \\
  Department of Electrical Engineering\\
  Indian Institute of Technology Jodhpur\\
  Jodhpur, India 342037\\
}

%
%%%%%%%%%%%%%%%%%%%%%%%%%%%%%%%%%%%%%%%%%%%%%%%%%%%%%%%%%%%%%%%%%%%%%%%%%%%%%%%%%%%%%%%%%%%%%%%%%%%%%%%%%%%%%

\maketitle % typeset the header of the contribution

\begin{abstract}
Melanoma is a life-threatening form of skin cancer when left undiagnosed at the early stages. Although there are more cases of non-melanoma cancer than melanoma cancer, melanoma cancer is more deadly. Early detection of melanoma is crucial for the timely diagnosis of melanoma cancer and prohibit its spread to distant body parts. Segmentation of skin lesion is a crucial step in the classification of melanoma cancer from the cancerous lesions in dermoscopic images. Manual segmentation of dermoscopic skin images is very time consuming and error-prone resulting in an urgent need for an intelligent and accurate algorithm. In this study, we propose a simple yet novel network-in-network convolution neural network(CNN) based approach for segmentation of the skin lesion. A Faster Region-based CNN (Faster RCNN) is used for preprocessing to predict bounding boxes of the lesions in the whole image which are subsequently cropped and fed into the segmentation network to obtain the lesion mask. The segmentation network is a combination of the UNet and Hourglass networks. We trained and evaluated our models on ISIC 2018 dataset and also cross-validated on PH\textsuperscript{2} and ISBI 2017 datasets. Our proposed method surpassed the state-of-the-art with Dice Similarity Coefficient of 0.915 and Accuracy 0.959 on ISIC 2018 dataset and Dice Similarity Coefficient of 0.947 and Accuracy 0.971 on ISBI 2017 dataset.
\end{abstract}

%%%%%%%%% BODY TEXT
\section{Introduction}
It is estimated by the Indian Cancer Society that 55,100 new cases of melanoma are being
diagnosed each year in 
India \cite{india}.
Worldwide, it caused more than 60,000 deaths out of the 350,000 cases reported in 2015 and causes one death in every 54 minutes in  US ~\cite{report}.
Even after being accounted for as less as 1\% of the total skin-related diseases, melanoma cancer has become the major cause of death in these diseases. The annual cost of healthcare for melanoma cancer exceeds \$8 billion~\cite{A_Cost}. With early detection of melanoma, the 5-year survival rate can be increased up to 99\%; whereas delaying the diagnosis can drastically reduce the survival rate to 23\%~\cite{death} once it spreads to other parts of the body. Hence, it is of fundamental importance to identify the cancerous skin lesion at the earliest to increase the survival rate.

A huge amount of time and effort has been dedicated to increasing the accuracy and scale of diagnostic methods by researchers worldwide. The International Skin Imaging Collaboration(ISIC) provides a publicly available dataset of more than 25,000 dermoscopy images. They have been hosting the benchmark competition on skin lesion analysis every year since 2016. The previous year challenge comprised of 3 tasks on lesion analysis: Lesion Boundary Segmentation, Lesion Attribute Detection, and Lesion Diagnosis. % 
The extraction of crucial information for accurate diagnosis and other clinical features depends heavily on the lesion segmented from the given dermoscopic image and therefore, segmentation of the lesion has been designated as a decisive prerequisite step in the diagnosis~\cite{rethink,survey}.

Our work majorly focuses on the segmentation of the skin lesions which in itself is a challenging task. Skin lesions are accompanied by a huge variance in shape, size, and texture. While melanomas have very fuzzy lesion boundaries, there are further artifacts introduced due to hair, contrast, light reflection, and medical gauze which makes it more difficult for the CNN-based approach to segment the lesions.

%-------------------------------------------------------------------------

\subsection{Related}

In recent years, many pixel-level techniques for skin lesion segmentation have been developed. Initial work explored the visual properties of skin lesion like color and texture and applied classical techniques. Li et al.~\cite{Depth} proposed the use of above-mentioned features with classical edge detection for a contour-based methodology. Garnavi et al. \cite{garnavi2009skin} combined use of histogram thresholding and CIE-XYZ color space to segement the lesion. While classical approaches do not generalize well on unseen lesion images, deep convolutional neural network(DCNN) based approaches have proven to be a great success in generalization over such tasks with improved accuracy and precision~\cite{unet,v-net}.
Mishra et al. \cite{deep} presented an efficient implementation of UNet \cite{unet} and compared the improvement in performance with other classical methods. Many authors have used the same CNN-based method for segmentation as well as classification. In~\cite{twofrcn}, two fully convolutional residual networks(FCRN) were used, to segment as well as classify skin lesions at the same time.

Instead of processing only the local features, much work has been focused on processing the global information with CNNs as well. In~\cite{multires}, the authors made use of pyramid pooling to incorporate global context along with spatial information to produce location precise masks. Some work has also introduced the concept of processing the features selectively. In~\cite{focusnet}, the authors made use of  Squeeze-and-Excitation~\cite{squeeze} network to incorporate attention for focusing only on the important parts of the feature maps.~\cite{gan2} and~\cite{denseres} made use of modified GANs(General Adversarial Network) which involves training a Generator and Discriminator network in an adversarial fashion to generate accurate segmentation maps. Recently, two-stage CNN pipelines have also been researched upon. In~\cite{grabcut}, they used YOLO \cite{yolo}, an object detection network for localising the skin lesion and then employed classical image segmentation algorithm for segmenting the lesion. Similarly,~\cite{shreyas} first employed Faster RCNN~\cite{frcnn} for detecting the lesion in the images and then employed a dilation-based autoencoder for segmenting the lesion. 

\section{Methods}
\begin{figure*}
    \centering
    \centering\includegraphics[height=1.6cm,width=12.2cm]{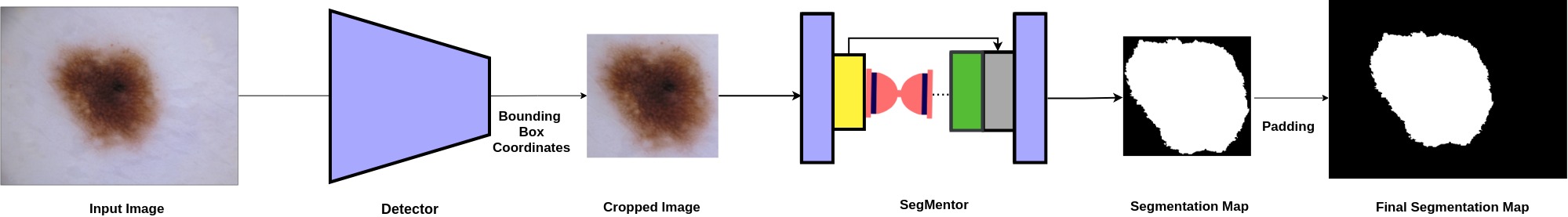}
    \caption{An overview of the proposed pipeline.}
    \label{fig:Overall}
\end{figure*}

We took inspiration from~\cite{grabcut} and~\cite{shreyas} and to apply the pre-processing step of extracting the region of interest(ROI). We implemented the faster region-based convolutional neural network (Faster RCNN)~\cite{frcnn} on the lines of \cite{shreyas} as Stage 1 of our pipeline. The ROIs extracted through our detection network were give as input to our segmentation network to obtain the lesion mask which served as the Stage 2 of our pipeline. We named the Stage 1 as the Detector and the Stage 2 as the SegMentor. Our segmentation network is inspired from the UNet \cite{unet} and the Hourglass \cite{hg} network. As previously experimented in \cite{grabcut} and \cite{shreyas}, the localized and cropped image of the lesion area by the detector in the given dermoscopic images, were used to train the network along with the cropped segmentation masks. Doing so, increased the over all performance of their segmentation algorithms. The reason behind this being that a prior removal of irrelevant features and other nearby pixels from the input images and present only relevant features at segmentation stage. This helps the segmentation network to achieve good and fast results.
We used Dice Similarity Coefficient(DSC) as the evaluation metric which is similar to the Jaccard Index~\cite{jaccard}, to tackle the issue of imbalanced lesion to background ratio. We discuss the architectural uniqueness and the non-conventional training strategy followed for our model, where components of the model were trained sequentially to attain the optimal value of parameters. We have also shown a comparison between model performances with and without employing Stage 1 i.e. melanoma localization in table \ref{tab:results2}.
The overall pipeline is summarized in figure \ref{fig:Overall}. We named the combined pipeline of localization and segmentation networks as Detector-SegMentor. 

\subsection{Dataset}
The ISIC challenge 2018~\cite{d1,d2} dataset was used for training. The ISIC 2018 challenge provided datasets for skin lesion segmentation task along with attribute detection and lesion diagnosis tasks. The data was collected from various institutions and clinics around the word. ISIC archive is the largest dermoscopic image library available publicly. The challenge provided a total of 2594 images with their corresponding ground truth masks. Image dimensions varied from 1022x767 to 6688x4439. Out of the available dataset, we utilised 2000 images solely for training purpose. To extend our training data, we performed conventional augmentations such as horizontal and vertical flips, rotation, shear and stretch, central cropping, and contrast shift. The final training dataset consisted of 30,000 images with their corresponding ground truth masks. Moreover to showcase the generalisability of our model, we also used ISBI 2017 \cite{d17} and PH\textsuperscript{2} \cite{ph2} datasets for validation.

\subsection{Proposed Detector-SegMentor}
Faster-RCNN network in Stage 1 task gave us the localised lesion. The Detector returns a set of coordinates corresponding to the input image which confines a lesion in it with a  certain probability. The lesion area is cropped from the original image with the help of the obtained coordinates from the detector. The cropped image is then either resized (if aspect ratio larger than 512x512) or padded with zeros (if aspect ratio smaller than 512{x}512) to obtain the image size of 512{x}512{x}3 to be fed into the SegMentor (Segmentation Network) to generate segmentation maps. 
\subsubsection{Stage - 1 : Detector} 
\begin{figure*}
    \centering
    \includegraphics[height=4.7cm,width=12.3cm]{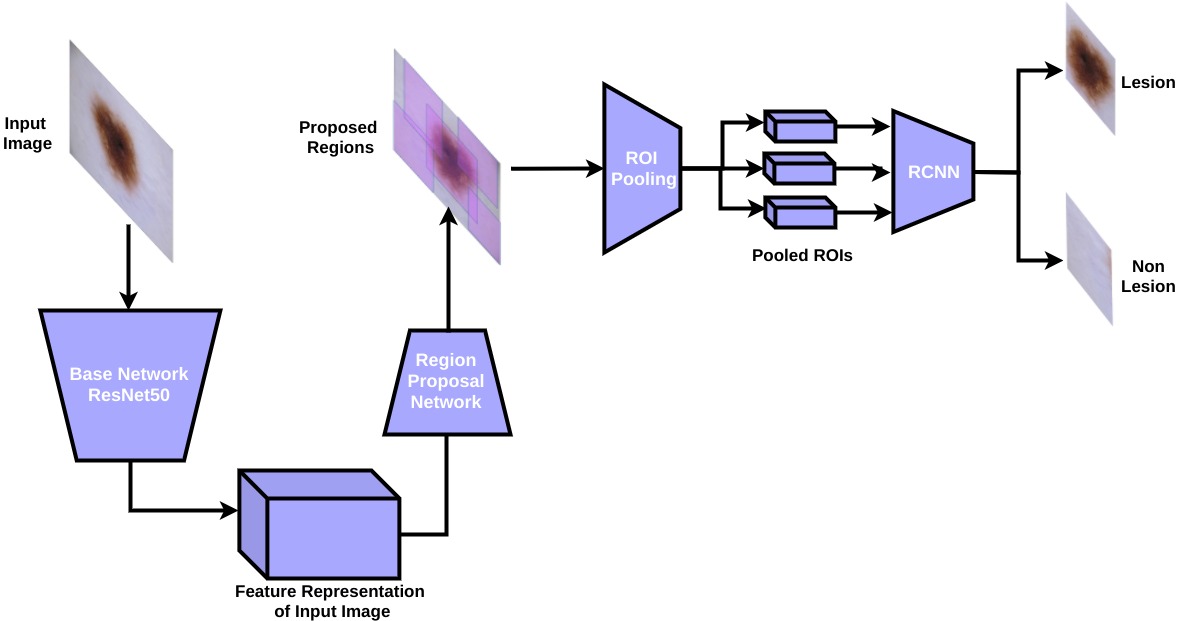}
    \caption{ Detector : Faster RCNN for skin lesion localization.}
    \label{fig:frcnn}
\end{figure*}

In Stage 1 of the proposed method, the skin lesion is localized in the input image and then passed on as input to the Stage 2. The Detector can be divided into three major components namely the base network, region proposal network(RPN), and the RCNN. 
The base network generates a feature map of the input image to be used by the RPN. We used the ResNet50 network \cite{resnet} and its pre-trained weights on ImageNet. The RPN then acts on the feature map from the base network and outputs the region proposals in the form of set anchor boxes. These boxes have a high chance of containing the lesion from the input feature map. Thereafter, each proposed region is classified into lesion/non-lesion and the bounding box coordinates of the proposal are trimmed to fit the lesion entirely by the RCNN. As given in \cite{shreyas}, time-distributed convolution layers were used for the RCNN to aid in avoiding repeated classification and also in accommodating the differing number of regions proposed by the RPN per image. Finally, Non-Maximum suppression was done with a threshold of 0.5 to remove the redundant boxes. Coordinates were scaled for the lesion in the original image. Figure \ref{fig:frcnn} depicts our Detector which comprises of three major components, namely the base network, RPN, and the RCNN.

\subsubsection{Stage - 2 : SegMentor}
After detecting the lesion at Stage 1, the detected section is cropped from the original image or padded appropriately to be given as input to the SegMentor. The segmentation network was designed in a network-in-network fashion. The base network has been derived from the well-known UNet~\cite{unet} which was a major breakthrough in biomedical image segmentation.

We proposed the use of a sequence of hourglass modules which are smaller but effectively dense networks at the bottleneck of the autoencoder. The module enabled us to further compress and better represent the bottleneck features for them to be easily decoded by the decoder. 

\begin{figure*}
    \centering
    \includegraphics[height=3.9cm,width=12.1cm]{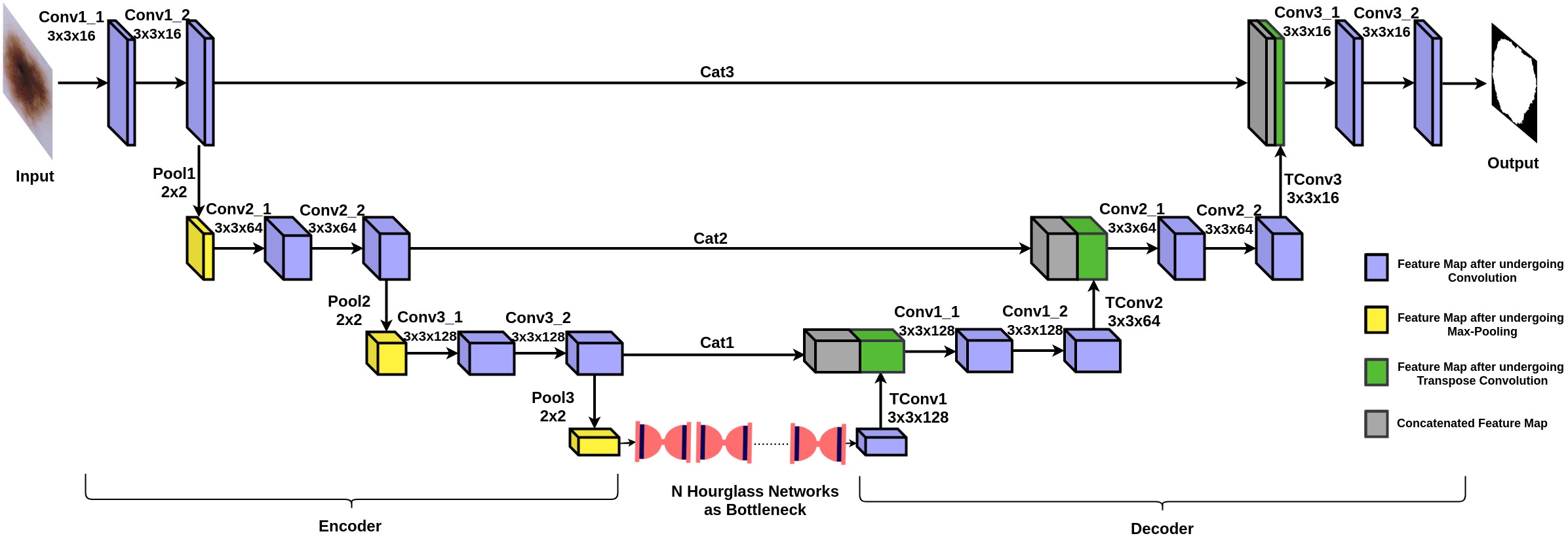}
    \caption { SegMentor : Encoder, Hourglass bottleneck and Decoder of the Segmentation network.}
    \label{fig:seg}
\end{figure*}

The encoder and decoder of the hourglass module are connected with processed skip connections instead of simple skip connections as in UNet \cite{unet}. We have demonstrated the effects of using multiple hourglass modules at the bottleneck in table \ref{tab:results2}. The results showed that the segmentation accuracy reached a maximum for the optimum number of modules. Increasing the modules beyond the optimum number resulted in overfitting of the network. The following paragraphs give an insight into the encoder, decoder, hourglass modules, and training strategy followed for the SegMentor network. The SegMentor is depicted in figure \ref{fig:seg}.

\textit {Encoder - Decoder : } The combination of an encoder and a decoder, termed autoencoder has been widely used across the literature for image-to-image translation tasks. Here, we exploited the same to obtain the mask for skin lesion. Our encoder comprises of convolution blocks where each block consist of 2 convolutional layers of filter size 3{x}3, each followed by batch normalization. We used max-pooling with steps 2{x}2 to downsample the features. Table 1 in supplementary material and figure \ref{fig:seg} describe the  detailed filter sizes and parameters for encoder. At the bottleneck of the encoder, feature map of size 64{x}64{x}128 was fed into the hourglass modules to obtain the compressed feature map of size  64{x}64{x}256 which was then fed into the decoder of the model. The decoder again comprises of  convolutional blocks each having 2 convolutional layer of filter size 3{x}3 and number of filter for different blocks are 16, 64 and 128 same as that of the encoder, detailed filter sizes and parameters are given in Table 4 in supplementary material. Both the encoder and decoder were connected using long skip connections. This facilitates better gradient flow through them and hence, tackles the issue of vanishing gradient in deep convolutional networks such as ours. The long skip connections between the encoder and decoder allow for the transfer of global features whereas, the hourglass modules provide local features in the form of compressed and better extracted feature maps to the decoder. This ultimately generates sharp and location-precise masks. The hourglass modules helped in tackling the variation in shape, size, and obstructions observed in the input images and made the model more robust to such variations as explained in the next paragraph.

\begin{figure*}
    \centering
    \centering\includegraphics[height=5.2cm,width=13.7cm]{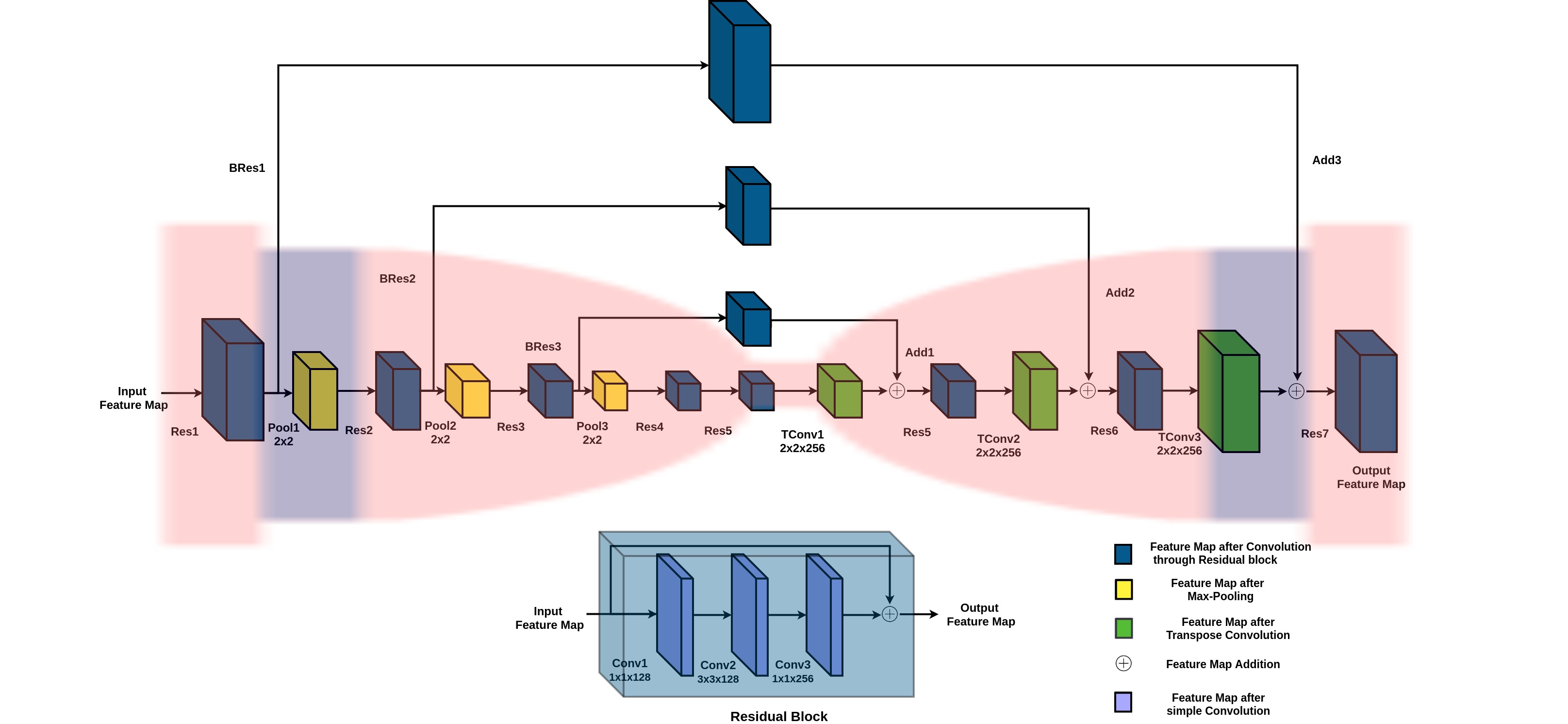}
    \caption{Hourglass Module and Residual Block.}
    \label{fig:hg}
\end{figure*}

\textit{Hourglass Module : } Hourglass modules are dense autoencoders placed at the bottleneck of the main encoder-decoder model. Hourglass modules have long as well as short skip connections, allowing for the better flow of information across the network. This leads to better extraction of the required feature map at output. The short skip connection is incorporated to make the network residual and hence, avoid gradient vanishing while also transfer the information at every step. Each residual block has 3 convolutional layers with batch normalization and a skip connection from input layer to output layer of each block as given in table 2 in supplementary material. Long skip connections between the hourglass-encoder and hourglass-decoder consist of intermediate residual blocks to process the skipped information before concatenating into the hourglass-decoder. This makes the network heavily dense allowing for better feature extraction and representation of bottleneck features. Table 3 in supplementary material and figure \ref{fig:hg} shows the complete architecture of the hourglass module.

\subsection{Training Strategy :}
The Detector was trained in step-wise manner where first, the RPN was trained followed by the regressor as described in~\cite{frcnn} and~\cite{shreyas}. Cross-entropy and categorical cross-entropy loss were used as classification loss in the RPN and RCNN respectively while the mean squared error (MSE) was used for both as regression loss. The masks obtained from the ISIC 2018 challenge were used to create synthetic ground truths for bounding box regression. Training strategy mentioned in~\cite{shreyas} was followed for Faster-RCNN which is an standard approach in general.

We used a step-wise training strategy for SegMentor where first, the autoencoder was trained for few epochs to learn representation of the data. Afterwards, a single hourglass module was introduced at the bottleneck of the encoder-decoder pair for training while keeping the weights of the encoder-decoder pair frozen. Second hourglass module was introduced following the first one for training where the weights of all other components were kept frozen for few epochs and finally, the entire model was trained freely. Multiple modules can be introduced with similar strategy. Table \ref{tab:results2} shows that as we increased the number of hourglass modules at the bottleneck of main network, the performance of framework increased till a certain point after which it started declining rapidly due to the overfitting on training dataset. Hence, we stopped at only 2 hourglass modules.

We used the dice coefficient loss for our segmentation model which is insensitive to class imbalance(poor foreground to background ratio).

\begin{figure*}
    \centering
    \includegraphics[height=2.2cm,width=5.2cm]{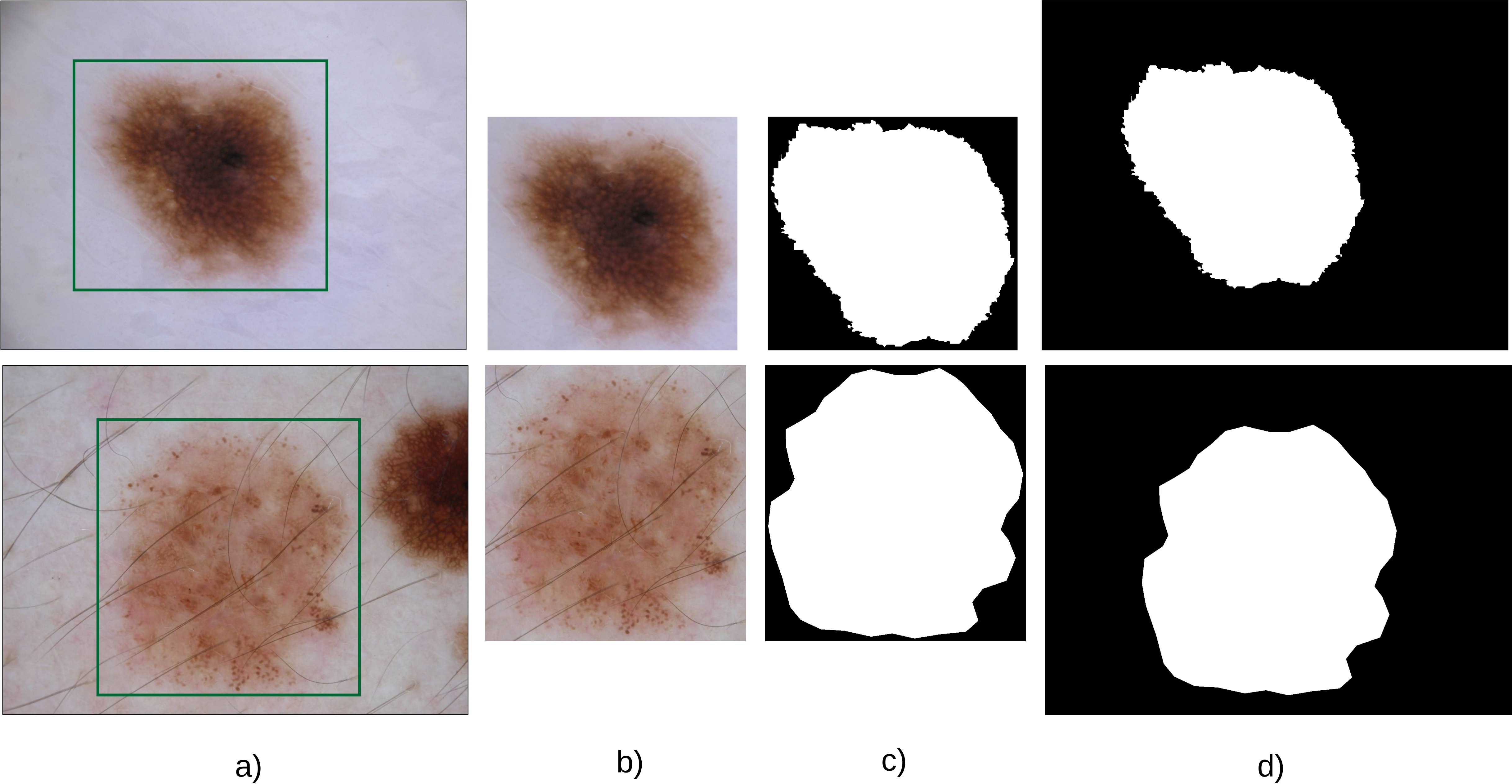}
    \caption  {Outputs at various stages of the pipeline. a) Lesion localized by the Detector, b) cropped image of the lesion, c) segmentation map from the SegMentor, d) final padded segmentation map.}
    \label{fig:data}
\end{figure*}

\section{Experimental Setup and Results}
We evaluated our framework on Dice Similarity Coefficient, Jaccard Index, Accuracy, Sensitivity and Specificity. We used the Adam optimizer with a learning rate of 0.00001 and 0.0002 for the Detector and SegMentor respectively. It roughly took 6 hours to train the Faster RCNN and 13 hours to train the SegMentor alone for 90 epochs for final end-to-end training with single hourglass module, both on NVIDIA's GTX 1080 Ti with 12 GB memory.

For evaluation, we randomly separated 594 images from ISIC 2018 challenge (training) dataset \cite{d1,d2} for pure testing purpose. 

Apart from ISIC 2018 \cite{d1,d2}, we evaluated our framework on PH\textsuperscript{2} \cite{ph2} and ISBI 2017 \cite{d17} datasets. 
The SegMentor was trained in multiple steps where initially only the encoder-decoder pair was trained for 20 epochs with a slightly smaller learning rate. Next, a single hourglass module introduced at the bottleneck and was trained alone for another 20 epochs with a slightly higher learning rate. Finally, the complete model was trained in an end-to-end manner for 90 epochs where the loss converged after 50 epochs. Tables~\ref{tab:results2} and ~\ref{tab:results} show the comparison among the results on ISIC 2018 \cite{d1,d2} and our results on ISBI 2017 \cite{d17} and PH\textsuperscript{2} \cite{ph2} as well. The supplementary material includes the qualitative (both success and failure) and quantitative (for ISBI 2017) results. Also, to further check the robustness of our proposed method, we performed 5-fold cross-validation on the ISIC 2018 dataset. The values obtained were 0.94(Accuracy), 0.903(Dice), and 0.783(Jaccard). From table 1, it can be seen that our method outperformed the famous MaskRCNN \cite{mask}. Though MaskRCNN also contains a detector and segmentor network, unlike ours, the training is done in an end-to-end fashion. In our method, the detector and segmentor are trained separately but resonate at the end. Also, MaskRCNN uses only simple 3x3 convolution layers to segment the lesion from the ROI but we have used a novel segmentation network altogether for the specified task.

\begin{table*}[]
\centering\caption{Comparison of results on ISBI  2017 validation set with UNet+2HG.}
\begin{tabular}{|c|c|c|c|c|c|}
\hline
\textbf{Method} & \textbf{Accuracy} & \textbf{Dice} & \textbf{Jaccard} & \textbf{Sensitivity} & \textbf{Specificity} \\ \hline
UNet \cite{unet}          & 0.920             & 0.768         & 0.651            & 0.853                & 0.957                \\ \hline
FocusNet \cite{focusnet}          & 0.921             & 0.831         & 0.756            & 0.767                & \textbf{0.989}               \\ \hline
Tu et al. \cite{denseres}          & 0.945             & 0.862         & 0.768            & 0.901               & 0.974                \\ \hline
Mishra et al. \cite{deep}          &0.928           & 0.868         & 0.842            & 0.930               & 0.954              \\ \hline
Proposed Method        & \textbf{0.971}             & \textbf{0.947}         &\textbf{0.844}           &\textbf{0.972}                & 0.981                \\ \hline
\end{tabular}
\label{tab:results11}
\end{table*}

\begin{table*}[]
\centering\caption{Results on ISIC 2018 validation set. In the table, UNet+nHG describes the network architecture used where 'n' represents the number of hourglass modules. Methods used to compare are taken from report \cite{d1} of the organisers of the ISIC 2018.}
\scalebox{0.85}{%
\begin{tabular}{|c|c|c|c|c|c|c|}
\hline
\textbf{Method}             & \textbf{Cropping Status} & \textbf{Accuracy} & \textbf{Dice}  & \textbf{Jaccard} & \textbf{Sensitivity} & \textbf{Specificity} \\ \hline
C. Qian (MaskRCNN)                    & -                        & 0.942          & 0.898        & 0.802       & 0.906             &0.963           \\ \hline
Y. Seok (Ensemble + C.R.F.)                     & -                        & 0.945            &0.904        &0.801         &0.934               &0.952         \\ \hline
Y. Ji (Feature Aggregation CNN)                      &-                        &0.943            & 0.900       & 0.799           & 0.964          &0.918              \\ \hline

Y. Xue (SegAN)                    & -                        & 0.945           & 0.903        & 0.798          & 0.940              &0.942              \\ \hline
\multicolumn{7}{c}{} \\ \hline
\multirow{2}{*}{UNet}       & W/o FRCNN                & 0.906             & 0.819          & 0.712            & 0.754                & 0.842                \\ \cline{2-7} 
                            & With FRCNN               & 0.917             & 0.85           & 0.746            & 0.842                & 0.891                \\ \hline
\multirow{2}{*}{UNet + HG}  & W/o FRCNN                & 0.928             & 0.841          & 0.746            & 0.821       & 0.924                \\ \cline{2-7} 
                            & With FRCNN               & 0.943             & 0.874          & 0.761            & 0.906                & 0.946               \\ \hline
\multirow{2}{*}{UNet + 2HG} & W/o FRCNN                & 0.937             & 0.887          & 0.773            & 0.912                & 0.944                \\ \cline{2-7} 
                            & With FRCNN               & \textbf{0.959}    & \textbf{0.915} & \textbf{0.809}   & \textbf{0.968}                &\textbf{0.973}               \\ \hline
\multirow{2}{*}{UNet + 3HG} & W/o FRCNN                & 0.921             & 0.866          & 0.756            & 0.893                & 0.931                \\ \cline{2-7} 
                            & With FRCNN               & 0.939             & 0.878          & 0.779            & 0.923                & 0.958                \\ \hline
\end{tabular}}
\label{tab:results2}
%\end{table*}

\vspace{6mm}
%\begin{table*}[]
\centering\caption{Results on ISBI 2017 validation and PH\textsuperscript{2} dataset with UNet+2HG.}
\scalebox{0.85}{%
\begin{tabular}{|c|c|c|c|c|c|}
\hline
\textbf{Dataset} & \textbf{Accuracy} & \textbf{Dice} & \textbf{Jaccard} & \textbf{Sensitivity} & \textbf{Specificity} \\ \hline
PH\textsuperscript{2}              & 0.979             & 0.952         & 0.891            & 0.975                & 0.988                \\ \hline
ISBI 2017        & 0.971             & 0.947         & 0.849            & 0.972                & 0.981               \\ \hline
ISIC 2018        & 0.959             & 0.915         & 0.809            & 0.968                & 0.973                \\ \hline
\end{tabular}
}
\label{tab:results}
\end{table*}

\begin{figure*}
    \centering
    \includegraphics[height=6cm,width=7cm]{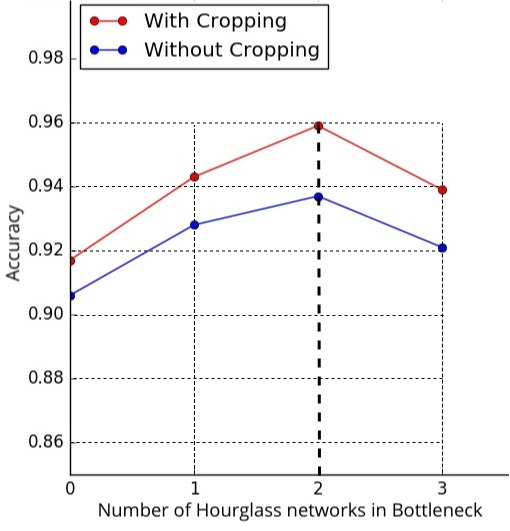}
    \caption{ Plot depicting variation of accuracy with the number of hourglasses in the bottleneck on ISIC 2018 validation set.}
    \label{fig:plot}
\end{figure*}

\begin{figure*}
    \centering
    \includegraphics[height=9.5cm,width=6cm]{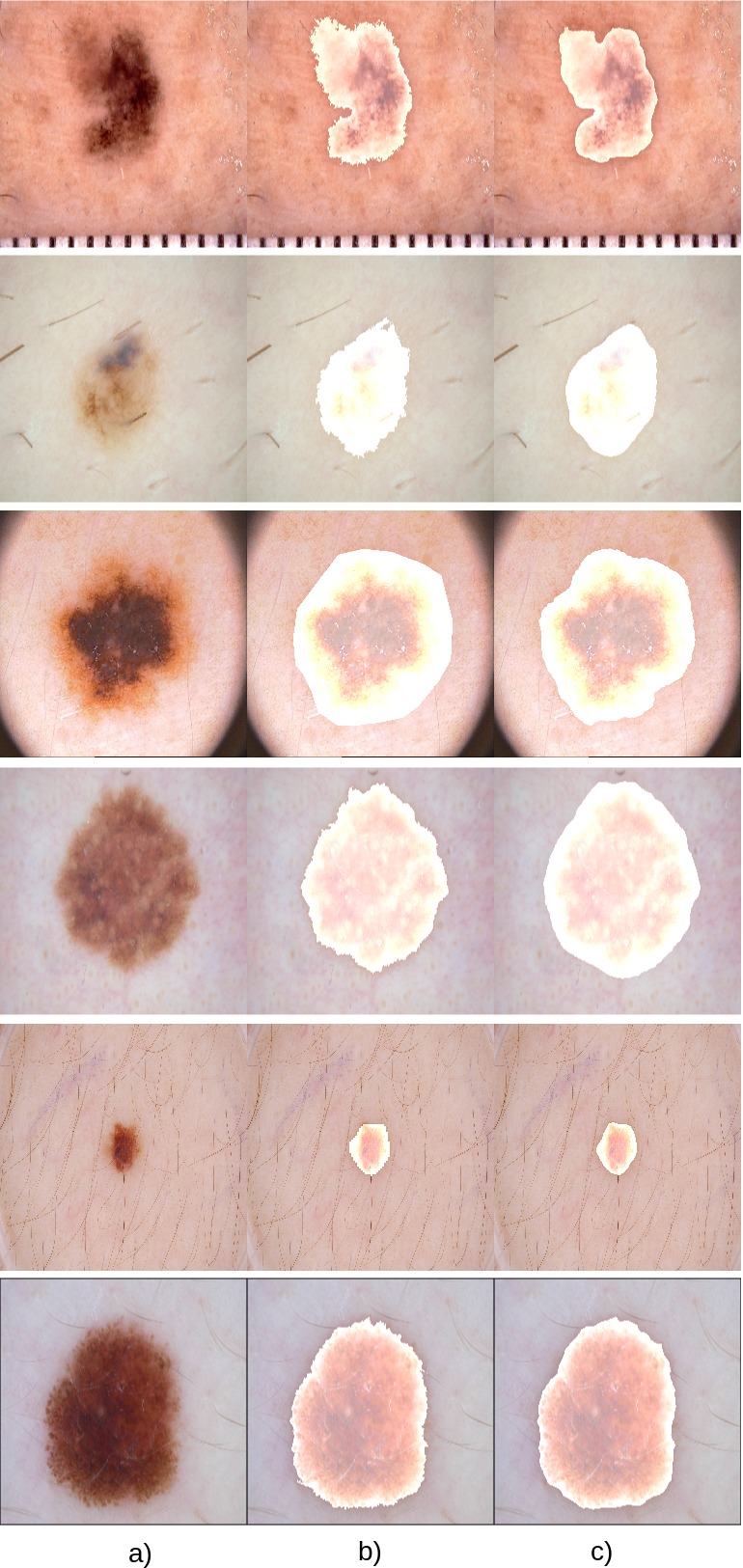}
    \caption{Qualitative results. a) Original images, b) ground truth lesion masks and c) predicted lesion masks.}
    \label{fig:examples}
\end{figure*}

We present most widely accepted comparative results on ISIC 2018 \cite{d1,d2} and performance on ISBI 2017 \cite{d17} and  PH\textsuperscript{2} \cite{ph2} datasets. While most of the literature have stated the comparative results on ISBI 2017 \cite{d17} and performance on  PH\textsuperscript{2} \cite{ph2} and in very few cases on ISIC 2018 \cite{d1,d2} dataset. ISIC 2018 \cite{d1,d2} dataset is somewhat the extension of the ISBI 2017 \cite{d17} dataset and our work majorly revolves around the 2018 dataset, we gave the comparative result on ISIC 2018 \cite{d1,d2} dataset only in main paper. To better evaluate and analyze our method we give comparative result on ISBI 2017 \cite{d17} dataset in table \ref{tab:results11} and compare our method with existing state-of-the-arts. Figure \ref{fig:examples} shows the qualitative results of our method.

\section{Conclusions}
The proposed methodology with multiple networks achieved the state-of-the-art on publicly available datasets namely ISIC 2018. The results show, confident lesion mask boundary obtained from our network. However, the results were less than the present state-of-the art in terms of specificity. This was mainly in cases where the contrast of the lesion matched with the contrast of the normal nearby skin and so some background was segmented as foreground. In future, we will try to improve the performance in terms of specificity while simultaneously also aim for one single end-to-end network architecture to perform the detection and segmentation task together. Also, it is planned to extend the generalization of the network to enable segmentation of skin lesion from images taken from normal mobile cameras.

\bibliographystyle{splncs04}
%\bibliography{main}

\end{document}